\documentstyle[twoside,fleqn,espcrc2,epsfig]{article}

\newcommand{\AmS}{{\protect\the\textfont2
  A\kern-.1667em\lower.5ex\hbox{M}\kern-.125emS}}

\newcommand{\lsim}{\mathrel{\mathop{\kern 0pt \rlap
  {\raise.2ex\hbox{$<$}}}
  \lower.9ex\hbox{\kern-.190em $\sim$}}}
\newcommand{\gsim}{\mathrel{\mathop{\kern 0pt \rlap
  {\raise.2ex\hbox{$>$}}}
  \lower.9ex\hbox{\kern-.190em $\sim$}}}

\begin{document}

\title{Sensitivity plots for WIMP modulation searches}

\author{S. Scopel\thanks{Attending speaker}, S. Cebri\'{a}n, E. Garc\'{\i}a,
         D. Gonz\'{a}lez, I. G. Irastorza, A. Morales, J. Morales, A. Ortiz de
         Sol\'{o}rzano, A. Peruzzi,
         J. Puimed\'{o}n, M. L. Sarsa,
         J. A. Villar\\
         \vspace{7 mm}
         Laboratorio de F\'{\i}sica Nuclear. Universidad de Zaragoza,
              50009, Zaragoza, SPAIN}

\begin{abstract}
Prospects of WIMP searches using the annual modulation signature
are discussed on statistical grounds, introducing
sensitivity plots for the WIMP--nucleon scalar cross section.
\end{abstract}

\maketitle

\section{INTRODUCTION}

The annual modulation effect\cite{drukier} provides a distinctive signature
for the identification of a Dark Matter signal in the direct searches
of WIMPS through their
elastic scattering off the nuclei of a detector.
Due to this effect the relative velocity between the detector and the
WIMP Mawellian distribution (assumed at rest in the Galactic rest
frame) is given by:
\begin{equation}\label{eq:earthvel}
  v_{\rm earth}=v_{\rm sun}+v_{\rm orb}
  \sin\delta\cos\left[\omega(t-t_0)\right]
\end{equation}
\noindent where
$v_{\rm sun}$ is the Sun's velocity in the galactic rest frame,
$v_{orb}\simeq$ 30 km sec$^{-1}$,
$\sin\delta\simeq$ 0.51 ($\delta$ is the angle between the Ecliptic
and the Galactic plane), $\omega=2\pi/T$, T=1 year and $t_0\simeq2^{nd}$ june.


\section{EXTRACTING THE MODULATION SIGNAL}

Given a set of experimental count rates $N_{ik}$ representing the number of events
collected in the i-th day and k-th energy bin, the mean value of $N_{ik}$
(expressed in number of counts per unit of detector mass, time and interval of recoil
energy) is:
\begin{eqnarray}\label{media}
  &&<N_{ik}>\equiv\mu_{ik}=\\
  &=&\left[b_{k}+S_{0,k}+S_{m,k}\cos\omega(t_i-t_0)\right]\cdot W_{ik}
\end{eqnarray}
\noindent where the $b_{k}$
represent the average background
while $S_{0,k}$ and  $S_{m,k}$ are the constant and the modulated amplitude
of the WIMP signal respectively. The
various parameters of the WIMP model are contained in $S_{0,k}$ and
$S_{m,k}$. In particular they depend on the WIMP-nucleus elastic cross sections
$\sigma$ and the WIMP mass $m_W$.
The $W_{ik}=M\Delta T_i\Delta E_k$ are the
corresponding exposures, where $M$ is the mass of the
detector, $\Delta E_k$ is the amplitude of the k-th energy--bin, while
$\Delta T_i$ represents the i-th time bin (in the following we will
assume all $\Delta T_i$= 1 day).
For simplicity $t_0$ will be omitted in the following equations.

The general procedure to compare theory with experiment is by
making use of the maximum-likelihood method.
The combined-probability
function of all the collected $N_{ik}$, assuming that they have a
poissonian distribution with mean values
$\mu_{ik}$, is given by:
\begin{equation}
  L=\prod_{ik}
  e^{-\mu_{ik}}\frac{\mu_{ik}^{N_{ik}}}{N_{ik}!}\label{likelyhood}.
\end{equation}

The most probable values of $m_W$ and $\sigma$ maximize
$L$ or, equivalently, minimize the function:
\begin{eqnarray}
&&  y(m_W,\sigma) \equiv -2 \log{L}-const  \label{ypiccolo} \\
                &=& 2 \mu -2\sum_{ik}N_{ik}\log
  \left[b_{k}+S_{0,k}+S_{m,k}\cos\omega t_i\right]\nonumber
\end{eqnarray}
\noindent where $\mu\equiv\sum_{ik}\mu_{ik}$ and
all the parts not depending on $m_W$ and $\sigma$  may be
absorbed in the constant because are irrelevant for the minimization.

\section{STATISTICAL SIGNIFICANCE OF THE SIGNAL}

Once a minimum of the likelihood function has been found, a positive result excludes
the absence
of modulation at some confidence level probability. This can be checked by
evaluating the quantity $\delta^2=y(\sigma=0)-y(m_W,\sigma)_{min}$ to test the
goodness of the null hypothesis.
In order to study the distribution of $\delta^2$ we make use of the asymptotic
behaviour:
\begin{eqnarray}\label{eq:asimptotic}
  &&\delta^2\simeq\chi^2(\sigma=0)-\chi^2_{min}\\
&&\chi^2(\sigma,m_W)\equiv \sum_k
 \frac{(S_{m,k}(m_W,\sigma)-X_{k})^2}{Var(X_{k})}\nonumber\\
&&X_k \equiv \frac{\sum_i N_{ik} \cos{\omega t_i}-N_{k}\beta_{k}}{W_k (\alpha_k
 -\beta_k^2)}.
\label{eq:xfreese}
\end{eqnarray}
\noindent where
$\beta_k\equiv \frac{\sum_i W_{ik} cos \omega
 t_i}{W_k}$, $\alpha_k\equiv \frac{\sum_i W_{ik} cos^2 \omega
 t_i}{W_k}$ and $N_k\equiv \sum_i N_{ik}$.
In the case of absence of a modulation effect
numerical simulations show that the quantity $\delta^2$ belongs asymptotically
to a $\chi^2$ distribution with two degrees of freedom.
We explain this
by the fact that once the cross section
$\sigma$ is set to zero the
likelihood function $L$ no longer depends on $m_W$ (all the $S_0$ and $S_m$ functions vanish)
and this is equivalent to fixing both the parameters of the fit at the same time.
In the case of presence of a modulation,
$\delta^2$ has the asymptotic distribution of a non central $\chi^2$ with one
degree of freedom and with a mean value given by
\begin{equation}\label{expansion1}
  <\delta^2>=\frac{1}{2}\sum_{k}
  \frac{S_{m,k}(\sigma,m_W)^2\Delta
  E_k}{b_{k}+S_{0,k}}MT\alpha+2\label{eq:magic}
\end{equation}
\noindent
where the same days of data taking have been assumed
for all the energy bins, and the
approximations $\sum_i N_{ik} \cos^2\omega t_i\simeq <N_{ik}>\sum\cos^2\omega
t_i$, $<N_{ik}>\simeq W_k (b_k+S_0)$ have been made.
In Eq.(\ref{eq:magic}) we have also defined the factor of merit
$\alpha\equiv \frac{2}{T}\sum_i \cos^2 \omega t_i$ ($\alpha$=1
in case of a full period of data taking)
and the terms depending on the $\beta_k$ have been neglected.

Since
the degree of overlapping between the distributions of $\delta^2$
in the two cases of absence and presence of modulation depends on
$<\delta^2>$,
equation (\ref{eq:magic}) allows to estimate the needed
exposure MT$\alpha$ in order to observe a modulation effect
with a given probability: for instance, $<\delta^2>$=14.9 (5.6)
corresponds to a 90\% (50\%) probability to see an effect at least at the 95\% (90\%) C.L.
Once a required $<\delta^2>$ is chosen, a sensitivity plot may be obtained by
showing the curves of constant MT$\alpha$ in the plane $m_W$--$\sigma$.



\section{SENSITIVITY PLOTS AND QUANTITATIVE DISCUSSION}

\begin{table*}[htb]
\setlength{\tabcolsep}{1.5pc}
\newlength{\digitwidth} \settowidth{\digitwidth}{\rm 0}
\catcode`?=\active \def?{\kern\digitwidth}
\caption{\small Summary of minimal exposures, all in kg $\cdot$ year.
Values off (in) parenthesis refer to $v_{\rm loc}$=220(170) km sec$^{-1}$.
E$_{th}$ indicates the energy
thresholds expressed in keV, b the background (assumed not dependent on energy) in
cpd/kg/keV.
Exposures are estimated for the WIMP mass range 10$\lsim
m_W\lsim$1000 unless specified otherwise.\label{tabella}}
\begin{tabular*}{\textwidth}{@{}l@{\extracolsep{\fill}}cccc}
\hline
                 & \multicolumn{1}{c}{Exploration of not}
                 & \multicolumn{1}{c}{DAMA region}
                 & \multicolumn{1}{c}{DAMA region}         \\
                 & \multicolumn{1}{c}{excluded regions}
                 & \multicolumn{1}{c}{$\delta^2=5.6$}
                 & \multicolumn{1}{c}{$\delta^2=15$}         \\
\hline
Ge, E$_{\rm th}=2$, b=$0.1$    & $80(50)$ & $50(25)$ & $175(90)$  \\
Ge, E$_{\rm th}=12$, b=$0.01$    & $25(19)^{*}$ & $50(25)$ & $190(95)$  \\
TeO$_2$, E$_{\rm th}=5$, b=$0.01$    & $40(25)$ & $40(20)$ & $150(80)$  \\
NaI, E$_{\rm th}=2$, b=$0.1$    & $50^{\dagger}$ & $180(100)$ & $660(355)$  \\
\hline
\multicolumn{5}{@{}p{120mm}}{$^{*}$ 45 GeV $\lsim m_W\lsim$110 GeV;
$^{\dagger}$m$_W\lsim$70(125) GeV.}

\end{tabular*}
\end{table*}

\begin{figure}
\begin{center}
\vspace{9pt}
\mbox{\psfig{figure=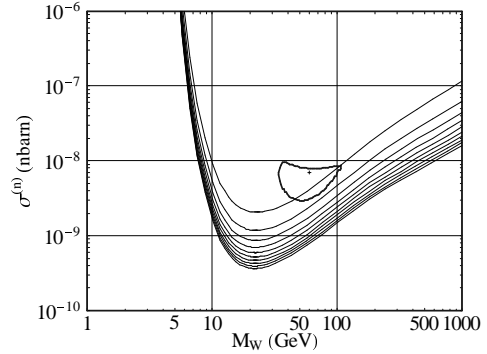,height=35mm}}
\end{center}
\caption{\small Sensitivity plots in the $\sigma^{(n)}$--$m_W$ plane for
a germanium detector.
\label{fig:ge_eth2_b001_mtg}}
\end{figure}

In Figures 1--2 we
discuss the example of a Germanium detector
with background b=0.01 cpd/kg/keV (assumed constant with energy) and
energy thresholds E$_{th}$=2 keV, values not unrealistic, taken into account the
recent performances of some Ge
detectors. Parameter values used in the plots are the local halo mass density
$\rho$=0.3 GeV/cm$^3$, $v_{\rm loc}$=220 km sec$^{-1}$ (
$v_{\rm loc}$ is the measured rotational velocity of
the Local System at the Earth's position),
the WIMP r.m.s. velocity $v_{\rm rms}^2=\frac{2}{3} \; v_{\rm loc}^2$
and $v_{\rm sun}\simeq (v_{\rm loc}+12)$ km sec$^{-1}$.

In Figure 1 the sensitivity plots
for $<\delta^2>$=5.6 is shown in the plane $m_W-\sigma^{(n)}$, where $\sigma^{(n)}$
is the WIMP cross section $\sigma$ rescaled to the nucleon by adopting a
scalar--type interaction.
The different curves correspond to
values of MT$\alpha$ from 10 kg$\cdot$ year to 100 kg$\cdot$ year
in steps of 10 (from top to bottom). The closed contour and the cross indicate
respectively
the 2$\sigma$ C.L. region singled out by the DAMA modulation search experiment and the minimum
of the likelihood function found by the same authors\cite{dama}.
Note that an exposure of 10 kg$\cdot$year of a Ge detector of the
above--quoted performances would explore almost totally the DAMA
region.

In Figure 2 we show, as a function of $m_W$, the minimal exposures required for the
same germanium set--up in order for its sensitivity contour to lie below the
upper limit on $\sigma^{(n)}$ implied by the exclusion plot
obtained in Ref.\cite{dama} with pulse shape discriminated
spectra.
Since the present uncertainty in $v_{\rm loc}$ can affect the results in a significant
way\cite{damabott}, the different values $v_{\rm loc}$= 170, 220 and
270 km sec$^{-1}$ are shown by the dotted, solid and dashed curves respectively. In each
case the values b=0,0.01 and 0.1 are given from bottom to top.

Some examples of minimal exposures for other target materials are given
in Table 1 for $v_{\rm loc}$= 220 and 170 km sec$^{-1}$.
The second column of table 1 shows the exposures
necessary to explore the regions of the $m_W$--$\sigma^{(n)}$ plane
below the exclusion plot or Ref.\cite{dama}. The third and fourth columns show the
lowest values of MT$\alpha$ that give a sensitivity plot encompassing
all the 2$\sigma$ DAMA contour for $<\delta^2>$=5.6 and 15 respectively.
The experimental thresholds and resolutions assumed in Table 1 are close to those already
obtained (or foreseeable) in Ge, NaI and TeO$_2$ detectors.

A systematic study of sensitivity plots (not shown here for lack of
space\cite{noi}) concludes that prospects of modulation
searches seem promising provided that the WIMP signal
is not far below present sensitivities,
the lowest values of
explorable $\sigma^{(n)}$ falling in most cases in the typical range of
few $\times 10^{-10}$ nbarn. An important feature of all the plots is that the
sensitivity to modulation is generally a decreasing function of
the WIMP mass, the highest sensitivities corresponding roughly
to the interval $10 \;{\rm GeV}\lsim m_W \lsim 130\;{\rm GeV}$, and
depending in a sensitive way on the value of the parameter $v_{\rm loc}$.

\section { Acknowledgements}
This work has been partially supported by the Spanish Agency of
Science and Technology (CICYT) under the grant AEN99--1033 and the
European Commission (DGXII) under contract ERB-FMRX-CT-98-0167.
One of us (S.S.) acknowledges the partial
support of 
INFN (Italy).

\begin{figure}
\begin{center}
\mbox{\psfig{figure=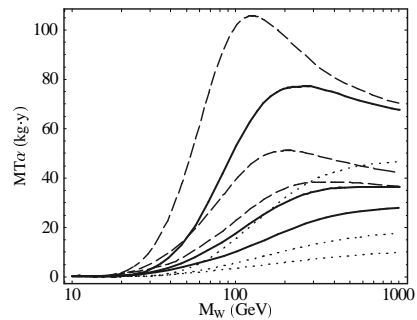,height=35mm}}
\end{center}
\caption{\small
Minimal exposures MT$\alpha$
for the $<\delta^2>$=5.6 calculated
for a germanium detector with threshold energy E$_{th}$=2 keV.}
\label{fig:ge_eth2_b001_mtmin}
\end{figure}


\begin{thebibliography}{9}
\footnotesize
\bibitem{drukier}
A. Drukier, K. Freese and D. Spergel, {\em Phys. Rev.} {\bf D33} (1986)
3495; K. Freese, J. Frieman and A. Gould, {\it Phys. Rev.} {\bf D37} (1988) 3388.


\bibitem{damabott}
P.Belli et al., {\em Phys. Rev.} {\bf D61} (2000) 023512; M.
Brhlik and L. Roszkowski, {\em Phys. Lett.} {\bf B464} (1999) 303.


\bibitem{dama}
R. Bernabei et al., {\em
  Phys. Lett.} {\bf B450} (1999) 448; {\em Phys. Lett.} {\bf B389} (1996) 757.

\bibitem{noi}
S. Cebri\'{a}n et al, hep-ph/9912394.


\end{thebibliography}
\end{document}